# Traffic Allocation for Low-Latency Multi-Hop Networks with Buffers

Guang Yang, *Student Member, IEEE*, Martin Haenggi, *Fellow, IEEE*, and Ming Xiao, *Senior Member, IEEE*

*Abstract*—For buffer-aided tandem networks consisting of relay nodes and multiple channels per hop, we consider two traffic allocation schemes, namely local allocation and global allocation, and investigate the end-to-end latency of a file transfer. We formulate the problem for generic multi-hop queuing systems and subsequently derive closed-form expressions of the end-to-end latency. We quantify the advantages of the global allocation scheme relative to its local allocation counterpart, and we conduct an asymptotic analysis on the performance gain when the number of channels in each hop increases to infinity. The traffic allocations and the analytical delay performance are validated through simulations. Furthermore, taking a specific two-hop network with millimeter-wave (mm-wave) as an example, we derive lower bounds on the average end-to-end latency, where Nakagami-$m$ fading is considered. Numerical results demonstrate that, compared to the local allocation scheme, the advantage of global allocation grows as the number of relay nodes increases, at the expense of higher complexity that linearly increases with the number of relay nodes. It is also demonstrated that a proper deployment of relay nodes in a linear mm-wave network plays an important role in reducing the average end-to-end latency, and the average latency decays as the mm-wave channels become more deterministic. These findings provide insights for designing multi-hop mm-wave networks with low end-to-end latency.

## I. INTRODUCTION

### A. Background and Motivation

In many future applications, low latency is a crucial quality-of-service (QoS) constraint [1]. For instance, vehicle-to-everything, remote surgery, and industrial control need the support of low-latency communications. Accordingly, the requirement on end-to-end latency in the fifth-generation (5G) mobile network and beyond, on the order of 1 to 5 ms, is much more stringent than that in 3G and 4G systems [2], [3]. Besides, to handle the unprecedented data volumes and heavy traffic load in future wireless communications, large buffers are usually used at transceivers. With these buffers, the data can be temporarily stored in a queue, until the corresponding service is available for its delivery. The queuing delay is defined as the waiting time of a file in the buffer or queue before being transmitted [4], [5]. For future wireless systems with buffers, the queuing delay becomes a key contributor to the overall latency, since the heavy network traffic may produce significant data backlog in buffers. Thus, one of the most effective ways for achieving lower latency is to reduce the queuing delay.

It is worth mentioning that, as one key enabler of high data-rate transmissions, millimeter-wave (mm-wave) technologies have raised extensive research interest and have been regarded as promising candidates in future mobile networks [6]–[8]. Motivated by the huge potential of using mm-wave in various scenarios, in this work, we restrict ourselves to mm-wave bands to investigate the traffic allocations over multi-hop networks. In this work, we consider a linear multi-hop network that consists of a source node, a destination node, and multiple buffer-aided relay nodes, where parallel channels in each hop are also assumed. This network architecture is promising for future mobile networks and motivated by the following two facts:

(i) Unlike wireless communications in sub-6 GHz bands, mm-wave radios used in future mobile systems encounter much more severe path loss, which may restrict the range of wireless communications. One solution to enlarge the range of mm-wave communications is to use relay nodes. With the multi-hop architecture, the distance between adjacent nodes is shortened, thereby mitigating the serious path loss in mm-wave bands.

(ii) The consideration of several parallel channels in each hop mainly stems from the application of distributed antenna systems (DAS) or remote radio heads (RRH). Note that sharp beams are generated by the dense antenna elements in mm-wave bands. With DAS or RRH, multiple channels can be established between communication nodes with negligible inter-channel interference. Multiple channels in good conditions can be selected via proper channel estimation and tracking techniques, thereby enabling higher performance for mm-wave communications.

It is important to investigate the end-to-end latency for networks with the aforementioned multi-hop multi-channel architecture. However, this system model is rarely studied, especially when buffers are incorporated at relay nodes.

### B. Related Works

In the past few years, numerous efforts have been devoted to the research on latency in multi-hop networks with buffers, and remarkable progress has been reported. In [9], the average end-to-end delay in random access multi-hop wireless ad hoc networks was studied, and the analytical results were

This work was supported by EU Marie Curie Project, QUICK, No. 612652, and Wireless@KTH Seed Project "Millimeter Wave for Ultra-Reliable Low-Latency Communications".

G. Yang and M. Xiao are with the Department of Information Science and Engineering (ISE), KTH Royal Institute of Technology, Stockholm, Sweden (Email: {gy,mingx}@kth.se).

M. Haenggi is with the Department of Electrical Engineering, University of Notre Dame, Notre Dame, IN, 46556 USA. (E-mail:mhaenggi@nd.edu).

discussed and compared with the well established information-theoretic results on scaling laws in ad hoc networks. For an opportunistic multi-hop cognitive radio network, the average end-to-end latency in the secondary network was studied in [10] by applying queuing theoretic techniques and a diffusion approximation. In [11], the queuing delay and medium access distribution over multi-hop personal area networks was investigated.

To reduce the end-to-end latency in multi-hop queuing systems, many works have focused on various aspects such as the routing, scheduling, and traffic control. Using back-pressure methods, algorithms and analysis were widely investigated in [12]–[15] for low-latency multi-hop wireless networks. For a two-hop half-duplex network with infinite buffers at both the source and the relay node, the problem of minimizing the average sum queue length under a half-duplex constraint was investigated in [16]. Some efforts for systems with interference incorporated have also been made in the past decade. In [17], several QoS routing problems, i.e., end-to-end loss rate, end-to-end average delay, and end-to-end delay distribution, in multi-hop wireless network were considered, where an exact tandem queuing model was established, and a decomposition approach for QoS routing was presented. Using a tuple-based multidimensional conflict graph model, a cross-layer framework was established in [18], in order to investigate the distributed scheduling and delay-aware routing in multi-hop multi-radio multi-channel networks. Considering a multi-hop system that consists of one source, one destination, and multiple relays, the end-to-end delay performance under a TDMA-ALOHA medium access control protocol, with interferers forming a Poisson point process, was investigated in [19], and insights regarding delay-minimizing joint medium access control/routing algorithms were provided for networks with randomly located nodes. In the recent work [20], a distributed flow allocation scheme was proposed for random access wireless multi-hop networks with multiple disjoint paths, aiming to maximize the average aggregate flow throughput and guarantee a bounded packet delay.

In spite of many significant achievements in multi-hop networks with buffers, e.g., [11], [17], [21], [22], the research on traffic allocation to achieve low-latency schemes is rather limited. In our recent work [23], two low-latency schemes for mm-wave communications, namely traffic dispersion and network densification, were investigated in the framework of network calculus and effective capacity. The analysis in [23] was performed for a given network setting, i.e., fixed transmission scheme, sum power budget, and arrival rate for the whole network, and bounding techniques were used to explore the potential for low-latency communications. However, traffic allocations for reducing the latency were not investigated, and the potential performance gain by optimized traffic allocations was not quantified.

*C. Objective and Contributions*

For multi-hop networks with multiple channels in each hop, an optimized traffic allocation scheme plays a crucial role when incorporating queues [24], since the traffic congestion at the relay nodes due to non-optimized allocation may produce long queues, resulting in larger end-to-end latency [5]. Conventionally, the method for studying low-latency networks is to transfer the objectives into network optimization problems, i.e., problems in [25]. However, these graph-based approaches do not apply to the scenarios with buffers. Therefore, it is necessary to investigate the latency optimization problem for buffered networks in a different way.

The main objective of our work is to develop an efficient traffic allocation scheme for mm-wave networks that minimize the end-to-end latency[1]. Specifically, we consider a linear multi-hop buffer-aided networks with multiple channels in each hop. The main contributions of our work are summarized as follows:

- Focusing on two traffic allocation schemes, namely, local allocation and global allocation, we calculate the end-to-end latency for delivering a fixed-length message from the source to the destination. Furthermore, we analytically compare these two allocation schemes through the relative performance gain and investigate the benefits of global allocation.
- For multi-hop buffered networks with multiple channels in each hop, we exploit the recursive nature of the global allocation scheme. Thus, there is no need to search for the optimal solution in an exhaustive manner. The recursive method introduced in this paper significantly simplifies the global minimization of the end-to-end latency and provides insights for analyzing tandem queuing systems. Besides, we give the overall computational complexity for performing local or global allocations.
- Following the recursive characterization for global allocation, we present the asymptotic relative performance gain when the number of per-hop channels goes to infinity. Furthermore, based on traffic allocation schemes that can be applied to generic multi-hop networks, we specifically consider a two-hop linear mm-wave network and investigate the average end-to-end latency, with Nakagami-$m$ fading incorporated. We derive lower bounds of the average end-to-end latency for two allocation schemes.

## II. System Model and Problem Formulation

*A. System Model*

We consider a multi-hop system, which consists of multiple buffer-aided relay nodes and multiple channels in each hop. The first-in-first-out (FIFO) rule applies to the queues at the buffer-aided relay nodes. As illustrated in Fig. 1, given $n$ tandem relay nodes, we label all nodes in reverse order, i.e., from the destination to the source, to simplify the notation in the following analysis. That is, the destination and the source are labeled as node 0 and node $n+1$, respectively. The hop between node $h$ and node $h+1$ is denoted by hop $h$, for all $h \in \{0\} \cup [n]$. In addition, we assume there are $m_h$ channels in hop $h$, and we denote by $C_{h,k}$ the capacity of the $k^{\text{th}}$ channel in hop $h$ for all $k \in [m_h]$. We define $[N] \triangleq \{1, 2, \ldots, N\}$ for any $N \in \mathbb{N}$.

---

[1]The traffic allocations discussed in this paper can be generalized to networks at other frequency bands, e.g., those used in 3G/4G systems, while mm-wave is just taken as an important example for study here.
2

Regarding the channels in each hop, as aforementioned, due to the negligible multi-path effect and the high directivity of mm-wave beams, the small-scale fading is very weak. We denote by $C_{h,k}$ the channel capacity. First we assume no fading for the investigation in Sec. III and Sec. IV, thereby producing results for general networks with fixed capacities over all channels. In the following Sec. V, Nakagami-$m$ fading is added to the model. This modeling preserves the actual main characteristics of mm-wave channels in practice, and also provides high tractability for the following analysis.

In this work, we assume that the traffic allocation that assigns the traffic to the individual channels is performed at the source and relay nodes. That is, the traffic arriving at one node is decomposed into several fractions according to the given allocation scheme, and those fractions are subsequently pushed onto the channels and delivered to the next node, where each fraction is partitioned again. For the traffic allocation with respect to channels in hop $h$, we define the vector $\underline{\alpha}_h \triangleq [\alpha_{h,1}, \alpha_{h,2}, \ldots, \alpha_{h,m_h}] \in \mathbb{R}_+^{m_h}$ for all $h \in \{0\} \cup [n]$, which is subject to the following constraint

$$\|\underline{\alpha}_h\|_1 \triangleq \sum_{i=1}^{m_h} \alpha_{h,i} = 1, \quad (1)$$

where $\|\cdot\|_1$ represents the 1-norm for vectors. The traffic allocation $\underline{\alpha}_h$ is determined by the capacities of outgoing channels, such that the incoming fraction can be accordingly chopped down and reallocated onto the respective outgoing channels. It is worth noting that the traffic allocation $\underline{\alpha}_h$ is performed at node $h + 1$.

We assume multiple parallel servers at the source node and the relay nodes (the number of servers equals the number of outgoing channels), such that fractions can be transmitted over the different channels at the same time. This model follows a special variant of general *fork-join* systems [26], [27] (with a synchronization constraint), where all tasks of a job start execution simultaneously, and the job is completed when the final task leaves the system. Relay nodes are full-duplex but only equipped with a single buffer such that the data reception and transmission can be performed at the same time and the received fractions leave the buffer one by one. At each relay node, one fraction is not served (chopped into smaller fractions and forwarded to the next node) until it is completely received. In future mobile networks, the capacity of mm-wave channels can reach multi-gigabits per second and the packet size is around several kilobits or megabits at most [28], and hence the latency is of the order of milliseconds or even smaller. In this sense, compared to the scheme that allows to receive and transmit a fraction simultaneously, the setting that a fraction has to be received entirely before it can be forwarded over the next hop definitely produces higher latency (but not significantly), while avoiding the potential interference induced by simultaneous transmission and reception. Furthermore, we assume that the file is infinitely divisible, i.e., any fraction can be divided into smaller pieces with arbitrarily small size. Aiming to investigate the limits of low latency, we assume infinite divisibility throughout our work for theoretical purposes. However, for communication systems in practice, there always exists an atomic unit for constituting files, such that a file cannot be infinitely divisible. Hence, the delay performance obtained in this work is degraded if the practical constraint of finite divisibility is incorporated. It is worth mentioning that the assumption of infinite divisibility does not imply a fluid-flow model (although it is one of the typical features), since there are two phases, i.e., fraction reception and fraction transmission, at any intermediate node, rather than following the continuous manner in fluid-flow models.

*B. Problem Formulation*

For the transmission from the source to the destination, we consider two traffic allocation schemes, namely, local traffic allocation and global traffic allocation. To simplify the exposition, $\mathcal{M}_{\text{local}}$ and $\mathcal{M}_{\text{global}}$ are used to denote the local and global allocation scheme, respectively, which are described as follows.

- $\mathcal{M}_{\text{local}}$: Node $h$ for all $h \in [n + 1]$ only has the capacity information of the channels in hop $h - 1$. The traffic allocation performed at node $h$ only optimizes the transmission over channels in hop $h - 1$. This scheme ensures that the latency in the local hop is minimized, but is oblivious to the traffic allocations in other hops.
- $\mathcal{M}_{\text{global}}$: Node $h$ for all $h \in [n + 1]$ has the entire capacity information of all channels from hop 0 to hop $h - 1$. The traffic allocation performed at node $h$ not only relies on channels in hop $h - 1$, but also relies on channels in the remaining hops, i.e., from hop 0 to hop $h-1$. This scheme minimizes the latency through $h$ hops

The definition of the end-to-end latency in our study is given as follows:

**Definition 1** (End-to-End Latency). *Given traffic allocations $\{\underline{\alpha}_h\}$, $h \in \{0\} \cup [n]$, for all hops, the end-to-end latency $\tau_n(\underline{\alpha}_n, \underline{\alpha}_{n-1}, \ldots, \underline{\alpha}_0)$ (also written as $\tau_n$ in the sequel for notional simplicity) of the tandem queuing system with $n$ relay nodes is defined as the time to deliver one fixed-length file of size 1 (without loss of generality) from the source to the destination[2], describing the time span from the moment the source starts transmission to the moment all fractions are received at the destination.*

The definition above indicates that the latency takes into account both the time of traversing the wireless channels (service time) and the time of queuing in buffers (waiting time) at the relays. For exposition, we consider a simple network as an example to illustrate the end-to-end latency and the difference between $\mathcal{M}_{\text{local}}$ and $\mathcal{M}_{\text{global}}$, where $n = 1$, $m_1 = 2$, and $m_0 = 1$. In this specific network, traffic allocation $\underline{\alpha}_1 \triangleq [\alpha_{1,1}, \alpha_{1,2}]$ is performed over hop 1, while there is no traffic allocation over hop 0, i.e., $\underline{\alpha}_0 = 1$. The service time for fraction $\alpha_{i,j}$ over the channel with capacity $C_{i,j}$ is characterized by their quotient, i.e., $\alpha_{i,j} C_{i,j}^{-1}$. It is worth noting that, due to the adoption of buffer at node 1, the arrival order of different fractions should also be taken into account. Hence, in addition to the service time, the potential waiting time for the latter incoming fraction also contributes to the end-to-end latency $\tau_2$. Then $\tau_2$ is obtained as

$$\tau_2\left(\underline{\alpha}_1, \underline{\alpha}_0\right) = \begin{cases} \max\left\{\begin{array}{l} C_{0,1}^{-1} + \alpha_{1,1}C_{1,1}^{-1}, \\ \alpha_{1,2}\left(C_{0,1}^{-1} + C_{1,2}^{-1}\right) \end{array}\right\}, & \frac{\alpha_{1,1}}{C_{1,1}} \leq \frac{\alpha_{1,2}}{C_{1,2}} \\ \max\left\{\begin{array}{l} C_{0,1}^{-1} + \alpha_{1,2}C_{1,2}^{-1}, \\ \alpha_{1,1}\left(C_{0,1}^{-1} + C_{1,1}^{-1}\right) \end{array}\right\}, & \text{otherwise.} \end{cases} \quad (2)$$

---
[2]For generality and notational simplicity, we do not assign units to the file size and channel capacities, i.e., they are all normalized. For a concrete network, the most suitable units can be chosen, e.g., the file size (and, in turn, the fractions) could be measured in MB, and the capacities in Mb/s.

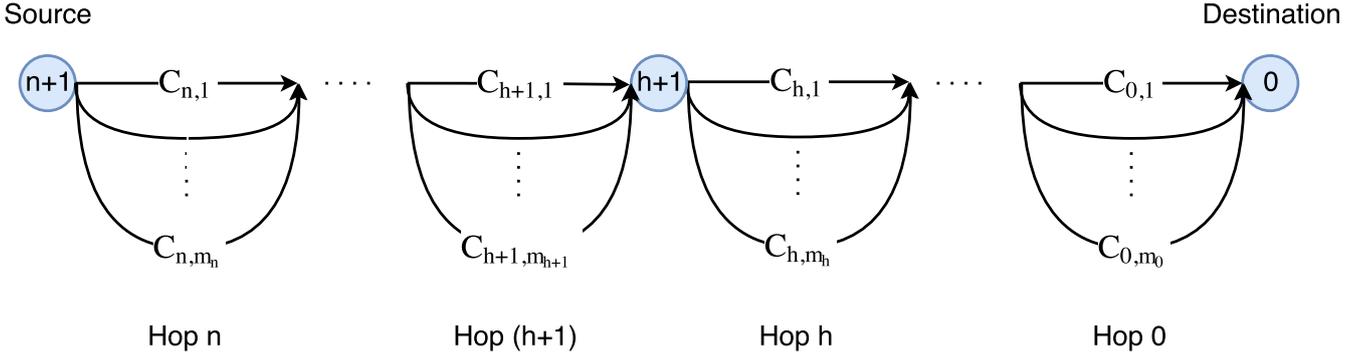

Fig. 1. Illustration of a multi-hop system, consisting of multiple relay nodes are multiple channels in each hop.

From (2), we see that the end-to-end latency considered in this work is different from those that consider either the service time or the waiting time only. Thus, the conventional techniques for graph-based network optimization or queuing systems are not applicable.

Since the traffic allocation only occurs at hop 1, the resulting minimization problem regarding $\tau_2$ can be formulated as

$$\tau_2^* = \min_{\|\underline{\alpha}_1\|=1} \tau_2\left(\underline{\alpha}_1, \underline{\alpha}_0\right) = \tau_2\left(\underline{\alpha}_1^*, \underline{\alpha}_0\right), \quad (3)$$

where $\underline{\alpha}_1^*$ representing the optimum traffic allocation differs for $\mathcal{M}_{\text{local}}$ and $\mathcal{M}_{\text{global}}$. Following the distinct mechanisms of $\mathcal{M}_{\text{local}}$ and $\mathcal{M}_{\text{global}}$, we have:

- for $\mathcal{M}_{\text{local}}$, since the channel information at the local hop is adopted for optimization, $\underline{\alpha}_1^*$ is obtained as

$$\underline{\alpha}_1^* = \arg\min_{\|\underline{\alpha}_1\|=1} \max\left\{\alpha_{1,1}C_{1,1}^{-1}, \alpha_{1,2}C_{1,2}^{-1}\right\}. \quad (4)$$

- for $\mathcal{M}_{\text{global}}$, since the channel information over all hops is adopted for optimization, $\underline{\alpha}_1^*$ is obtained as

$$\underline{\alpha}_1^* = \begin{cases} \arg\min_{\|\underline{\alpha}_1\|=1} \max\left\{\begin{array}{l} C_{0,1}^{-1} + \alpha_{1,1}C_{1,1}^{-1}, \\ \alpha_{1,2}\left(C_{0,1}^{-1} + C_{1,2}^{-1}\right) \end{array}\right\}, & \frac{\alpha_{1,1}}{C_{1,1}} \leq \frac{\alpha_{1,2}}{C_{1,2}} \\ \arg\min_{\|\underline{\alpha}_1\|=1} \max\left\{\begin{array}{l} C_{0,1}^{-1} + \alpha_{1,2}C_{1,2}^{-1}, \\ \alpha_{1,1}\left(C_{0,1}^{-1} + C_{1,1}^{-1}\right) \end{array}\right\}, & \text{otherwise.} \end{cases} \quad (5)$$

Evidently, $\mathcal{M}_{\text{global}}$ targets to the objective function $\tau_2$ straightforwardly for optimization, while $\mathcal{M}_{\text{local}}$ can only give a suboptimal solution via meeting the local optimization constraint.

It is worth noting that, when more hops and/or parallel channels on each hop are incorporated, it is not possible to provide a closed-form expression for the end-to-end latency (as given in (2)), since the orders of arrival of the fractions at different nodes become rather complicated.

## III. Traffic Allocation for Multi-Hop Networks

In this section, we will investigate the optimized end-to-end latency. In the analysis, the exact traffic allocations at the source and all relay nodes are presented, and the overall computational complexity for different traffic allocation schemes are briefly discussed subsequently. For generality, the capacities of parallel channels on each hop are distinguished by distinct notations. For specific scenarios considering the sum-capacity constraint, the bandwidth can be partitioned in any manner to implement parallel channels with arbitrarily distributed capacities.

We use "big-$O$" notation to characterize the overall computational complexity for $\mathcal{M}_{\text{local}}$ and $\mathcal{M}_{\text{global}}$. The definition of the "big-$O$" notation is given as follows: assuming $u(x)$ and $f(x)$ are functions defined on some subset $X \subset \mathbb{R}$, $O(f(x))$ denotes the set of all functions $u(x)$ such that $|u(x)/f(x)|$ stays bounded, i.e.,

$$O(f(x)) \triangleq \left\{u(x) : \sup_{x \in X} |u(x)/f(x)| < \infty\right\}. \quad (6)$$

Clearly, we have $O(f_1(x)) \subset O(f_2(x))$ if we have $|f_1(x)| \leq |f_2(x)|$ over all $x \in X$.

### A. Latency for Networks Using $\mathcal{M}_{\text{local}}$

Before deriving the minimum latency with $\mathcal{M}_{\text{local}}$, we start from a single-hop system, as shown in Fig. 2. To simplify the notation, we assume that there are $m$ channels between the source and the destination, where $C_i$ for $i \in [m]$ denotes the capacity of the $i^{\text{th}}$ channel. The traffic allocation $\underline{\alpha} \triangleq [\alpha_1, \alpha_2, \ldots, \alpha_m]$ is performed at the source node.

In the following Lemma 1, the optimal traffic allocation at the source node is presented, and the resulting minimum end-to-end latency for the system shown in Fig. 2 is also derived.

**Lemma 1.** *Given $m$ channels with capacity $C_i$ for $i \in [m]$ between the source and the destination, letting $\alpha_i \in (0, 1)$*

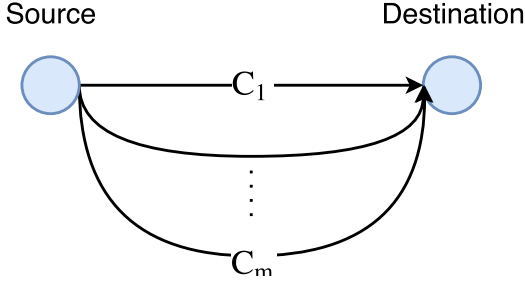

Fig. 2. Illustration of a single-hop system with multiple channels between the source and the destination.

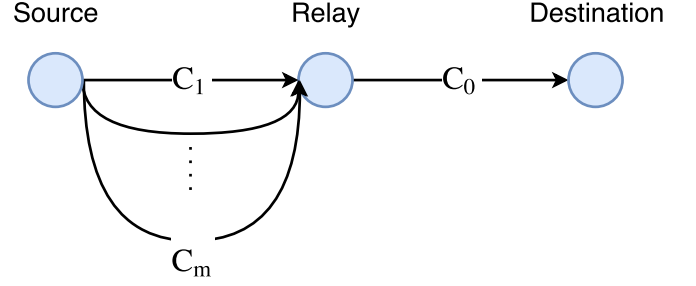

Fig. 3. Illustration of a two-hop network, where multiple channels are between the source and the relay, and one channel between the relay to the destination.

denote the fraction of the traffic allocated to the $i^{\text{th}}$ channel with capacity $C_i$, the minimum end-to-end latency is

$$\tau^* = \left(\sum_{i=1}^{m} C_i\right)^{-1}, \quad (7)$$

achieved by $\alpha_i = C_i \left(\sum_{j=1}^{m} C_j\right)^{-1}$ for all $i \in [m]$.

*Proof.* According to the mechanism of $\mathcal{M}_{\text{local}}$, we know that the minimum latency comes from applying the optimal traffic allocation $\underline{\alpha}^*$, i.e.,

$$\underline{\alpha}^* = \arg\min_{\|\underline{\alpha}\|=1} \max_{1 \leq i \leq m} \{\alpha_i C_i^{-1}\}, \quad (8)$$

which is solved as $\alpha_i = C_i \left(\sum_{j=1}^{m} C_j\right)^{-1}$ for all $i \in [m]$. Then the minimum end-to-end latency $\tau^*$ can be obtained by applying $\underline{\alpha}^*$. □

From Lemma 1, we notice that $\mathcal{M}_{\text{local}} = \mathcal{M}_{\text{global}}$ optimizes the traffic allocations such that all fractions arrive at the destination at the same time. Besides, from the perspective of the end-to-end delay, it is equivalent to having a single channel with the sum capacity when applying $\mathcal{M}_{\text{local}}$ for multiple channels.

Based on Lemma 1, the optimal local traffic allocation and the resulting minimum latency for the multi-hop system in Fig. 1 are obtained in Theorem 1.

**Theorem 1.** *For the tandem network in Fig. 1 with $n$ relay nodes and $m_h$ channels in the $h^{\text{th}}$ hop, the minimum end-to-end latency with $\mathcal{M}_{\text{local}}$ is*

$$\tau_n^* = \sum_{h=0}^{n} \left(\sum_{k=1}^{m_h} C_{h,k}\right)^{-1}, \quad (9)$$

*achieved by $\alpha_{h,k} = C_{h,k} \left(\sum_{j=1}^{m_h} C_{h,j}\right)^{-1}$.*

*Proof.* With $\mathcal{M}_{\text{local}}$, the objective is to ensure that all fractions can reach the next adjacent node simultaneously. According to Lemma 1, in hop $h$ the minimum latency is $w_h^* \triangleq \left(\sum_{k=1}^{m_h} C_{h,k}\right)^{-1}$.

Note that the traffic allocation is performed independently and sequentially from hop $n$ to hop $0$. In this case, the minimum end-to-end latency can be obtained by summing up $w_h^*$ over all $h \in \{0\} \cup [n]$, i.e., $\tau_n^* \triangleq \sum_{h=0}^{n} w_h^* =$ $\sum_{h=0}^{n} \left(\sum_{k=1}^{m_h} C_{h,k}\right)^{-1}$, when the proper local allocation scheme is applied. □

According to Theorem 1, it is not difficult to find that the end-to-end transmission with $\mathcal{M}_{\text{local}}$ is equivalent to transmitting the entire file hop by hop. Paired with Lemma 1, we know that all fractions are delivered from one node to the next simultaneously, which is equivalent to combining all sub-channels in each hop as a single channel with the sum capacity and moving the entire file through the network without partitioning. In this case, one relay node will buffer the whole file, while the buffers at other relay nodes remain empty. In terms of efficiency, the utilization of buffers for the transmission is relatively low, due to the uneven distribution of file fractions in the network. Furthermore, for networks with buffers, it is worth noting that the end-to-end latency (see the example with respect to (2)) differs from the non-buffered system, while the specific local scheme makes the resulting delay the same with the non-buffered system since arrivals on each hop can reach the node at the same time when $\mathcal{M}_{\text{local}}$ is applied.

### B. Latency for Networks Using $\mathcal{M}_{\text{global}}$

Prior to investigating the optimized end-to-end latency with $\mathcal{M}_{\text{global}}$ for the multi-hop network shown in Fig. 1, we consider a two-hop system shown in Fig. 3, where a buffer-aided relay node is deployed between the source and the destination. We assume $m$ channels between the source and the relay node, and we denote by $C_i$ for $i \in [m]$ the capacity of the $i^{\text{th}}$ channel. From the relay node to the destination, we assume that there is only one channel with capacity $C_0$. The traffic allocation $\underline{\alpha} \triangleq [\alpha_1, \ldots, \alpha_m]$ is performed at the source node, while no allocation is performed at the relay node, due to the single channel between the relay node and the destination.

In the following Lemma 2, the optimal traffic allocation at the source node and the resulting minimum end-to-end latency for the system shown in Fig. 3 are derived.

**Lemma 2.** *For the two-hop system shown in Fig. 3, the minimum end-to-end latency is*

$$\tau^* = \frac{(C_m^{-1} + C_0^{-1}) \prod_{k=1}^{m-1} C_{k+1} \left(C_k^{-1} + C_0^{-1}\right)}{1 + \sum_{i=2}^{m} \prod_{k=1}^{i-1} C_{k+1} \left(C_k^{-1} + C_0^{-1}\right)}, \quad (10)$$

*achieved by*

$$\alpha_i = \frac{\prod_{k=1}^{i-1} C_{k+1} \left( C_k^{-1} + C_0^{-1} \right)}{1 + \sum_{j=2}^{m} \prod_{k=1}^{j-1} C_{k+1} \left( C_k^{-1} + C_0^{-1} \right)}. \quad (11)$$

*Proof.* Please see Appendix A. □

Clearly, when applying the allocation in Lemma 2, one fraction reaches the relay node always at the time when the previous fraction has completely left the buffer. In contrast to the scheme in Lemma 1, all fractions arrive at the relay node sequentially, rather than simultaneously. Thus, the length of the resulting queue is the length of file fraction, which is usually much smaller than that of the whole file. It is worth mentioning that the optimum traffic allocation given in Lemma 2 is not unique, i.e., there exist several other solutions that achieve the same minimum end-to-end latency.

*1) $\mathcal{M}_{\text{global}}$ for a two-hop network:* Assuming the sum capacity for channels between the source and the relay is fixed, we next investigate the impact of increasing the number of channels between the source and the relay on end-to-end latency for the two-hop network in Fig. 3.

Given a fixed sum capacity, it is evident that $\tau^*$ depends on the individual capacities $C_i$ for $i \in [m]$. We will study the best channel capacity allocation for minimizing $\tau^*$ in the following corollary. Note that the channel capacity allocation determines the channel capacities potentially via frequency division techniques, while the traffic allocation partitions the data traffic according to the given channel capacities. We assume that the sum capacity for channels between the source and the relay is 1 without loss of generality.

**Corollary 1.** *For the two-hop system in Fig. 3, given $\sum_{i=1}^{m} C_i = 1$ with $C_i > 0$ for $i \in [m]$, the channel allocation that minimizes $\tau^*$ is the uniform allocation, i.e., $C_i = m^{-1}$ for all $i \in [m]$.*

*Proof.* Please see Appendix B. □

Given a sum-capacity constraint, Corollary 1 indicates that the lowest end-to-end latency with $\mathcal{M}_{\text{global}}$ can be achieved via the uniform allocation, i.e., via splitting the bandwidth equally.

Based on Corollary 1, the minimum end-to-end latency with $C_i = m^{-1}$ for any $i \in [m]$ is

$$\tau^* = \left( C_0 \left( 1 - \left( 1 + (mC_0)^{-1} \right)^{-m} \right) \right)^{-1}, \quad (12)$$

*achieved by*

$$\alpha_i = \frac{\left( 1 + (mC_0)^{-1} \right)^{i-1} (mC_0)^{-1}}{\left( 1 + (mC_0)^{-1} \right)^m - 1}. \quad (13)$$

We next investigate the monotonicity of $\tau^*$ in (12) and the asymptotic performance as $m \to \infty$.

**Corollary 2.** *For the two-hop system in Fig. 3, assuming $C_i = m^{-1}$ for all $i \in [m]$, $\tau^*$ monotonically decreases with $m$ and $\tau^* \to \left( C_0 \left( 1 - \exp\left( -C_0^{-1} \right) \right) \right)^{-1}$, when $m \to \infty$, and the limiting traffic allocation tends to be the uniform allocation.*

*Proof.* The monotonic decrease of $\tau^*$ with respect to $m$ follows since $(1 + x^{-1})^x$ increases with $x$. For the asymptotic latency, i.e., as $m \to \infty$, we have

$$\lim_{m \to \infty} \tau^* = \lim_{m \to \infty} C_0^{-1} \left( 1 - \left( 1 + (mC_0)^{-1} \right)^{-m} \right)^{-1}$$

$$= C_0^{-1} \lim_{m \to \infty} \left( 1 - \left( \left( 1 + (mC_0)^{-1} \right)^{mC_0} \right)^{-\frac{1}{C_0}} \right)^{-1}$$

$$= C_0^{-1} \left( 1 - \left( \lim_{m \to \infty} \left( 1 + (mC_0)^{-1} \right)^{mC_0} \right)^{-\frac{1}{C_0}} \right)^{-1}$$

$$= \left( C_0 \left( 1 - \exp\left( -C_0^{-1} \right) \right) \right)^{-1}. \quad (14)$$

With (13), we notice that

$$\lim_{m \to \infty} \frac{\alpha_{i+1}}{\alpha_i} = \lim_{m \to \infty} \left( 1 + (mC_0)^{-1} \right) = 1 \quad (15)$$

for all $i \in [m]$, which indicates that the limiting traffic allocation reduces to the uniform allocation. □

With a fixed sum capacity for channels between the source and the relay, it is demonstrated from Corollary 2 that it is always beneficial to increase the number of channels. This observation coincides with the well-known fact that dividing the bandwidth into as many fractions as possible is delay-optimum for full-duplex channels (see results, e.g., in [29]). However, this advantage diminishes as $m$ grows, and the resulting end-to-end latency with $\mathcal{M}_{\text{global}}$ approaches a constant limit, which only depends on $C_0$. This corollary indicates that, with respect to a given sum capacity, it is better to have multiple channels with smaller capacity rather than one single channel with larger capacity, or to split the channel into sub-channels using frequency division. Also, it is worth noting that the asymptotic result is a lower bound on the latency for all cases where the sum capacity over the first hop is fixed, and the lower bound is determined by $C_0$ when applying $\mathcal{M}_{\text{global}}$.

*2) Comparison with time division:* In what follows, we consider a scenario that has only a single channel with capacity 1 (rather than multiple sub-channels shown in Fig. 2) in the source-relay hop (corresponding to the systems that have only a single server at the source node). The file is partitioned into fractions via the traffic allocation in the time domain, which are sequentially delivered from the source node. Following the method for the proof of Lemma 2, the minimum latency can be achieved if the condition $\alpha_i = C_0^{-1} \alpha_{i-1}$ holds for all $i \in [m] \setminus \{1\}$. The performance for traffic allocation in the time domain with $\mathcal{M}_{\text{global}}$ is given as below.

**Lemma 3.** *Using time division, the minimum end-to-end latency $\tau^*$ with $\mathcal{M}_{\text{local}}$ is*

$$\tau^* = C_0^{-1} \frac{\sum_{i=0}^{m} C_0^i}{\sum_{i=0}^{m-1} C_0^i} = \frac{1 - C_0^{m+1}}{C_0 - C_0^{m+1}}, \quad (16)$$

*achieved by*

$$\alpha_i = C_0^{1-i} \sum_{k=0}^{m-1} C_0^{-k}. \quad (17)$$

$\tau^*$ *monotonically decreases with $m$, and $\tau^* \to \max \left\{ C_0^{-1}, 1 \right\}$ when $m \to \infty$.*

*Proof.* The detailed derivation for (16) is omitted since the method is similar to that in Appendix A. The monotonic decease of $\tau^*$ as $m$ increases is evident by observing (16). For the limiting latency as $m \to \infty$, we have

$$\lim_{m \to \infty} \tau^* = C_0^{-1} \lim_{m \to \infty} \frac{\sum_{i=0}^{m} C_0^i}{\sum_{i=0}^{m-1} C_0^i} = \begin{cases} 1, & C_0 \geq 1 \\ C_0^{-1}, & C_0 < 1, \end{cases} \quad (18)$$

which can be summarized as $\lim_{m \to \infty} \tau^* = \max\{C_0^{-1}, 1\}$. Therefore, the proof is completed. $\square$

We can see that the bottleneck channel (with smaller capacity) in the two-hop system determines the limiting latency. Unlike Corollary 2, the file fractions are not simultaneously delivered from the source in Lemma 3. This traffic allocation is performed using time division techniques. It is worth mentioning that we aim to study the minimum latency among all feasible traffic allocations in the time domain. We find that the optimal traffic allocation (with $\mathcal{M}_{\text{global}}$) based on the time division follows a geometric progression with the scale factor $C_0^{-1}$, while the uniform traffic allocation (with $\mathcal{M}_{\text{local}}$), i.e., $\alpha_i = m^{-1}$ for $i \in [m]$, cannot achieve the optimum.

For notational simplicity, we denote by $\tau_f^*$ and $\tau_t^*$ the minimum latency in (12) and (16), corresponding to the traffic allocations in the frequency domain and the time domain, respectively. We compare $\tau_f^*$ and $\tau_t^*$ for any positive integer $m$ in the following corollary.

**Corollary 3.** $\tau_f^* \geq \tau_t^*$ holds for all $m \in \mathbb{N}$.

*Proof.* Note that for $C_0 > 0$ we have

$$\left(C_0 + m^{-1}\right)^m = \sum_{i=0}^{m} m^{-i} \binom{m}{i} C_0^{m-i} \leq \sum_{i=0}^{m} C_0^i, \quad (19)$$

where the property $\binom{m}{i} \leq m^i$ for all $i \in \{0\} \cup [m]$ is applied, and the equality holds if $m = 1$ or $i = 0, 1$. Then we can obtain that

$$\left(1 + (mC_0)^{-1}\right)^{-m} \geq \frac{C_0^m}{\sum_{i=0}^{m} C_0^i}, \quad (20)$$

which leads to

$$\left(1 - \left(1 + (mC_0)^{-1}\right)^{-m}\right)^{-1} \geq \frac{\sum_{i=0}^{m} C_0^i}{\sum_{i=0}^{m-1} C_0^i}, \quad (21)$$

thereby concluding $\tau_f^* \geq \tau_t^*$. $\square$

Corollary 3 demonstrates that, if the total channel capacity in the source-relay hop is fixed, the time-division traffic allocation outperforms the frequency-division scheme in achieving the lower end-to-end latency.

*3) $\mathcal{M}_{\text{global}}$ for general network:* Based on Lemma 2, for the multi-hop buffer-aided network illustrated in Fig. 1, the optimal traffic allocation at the source node and the resulting minimum end-to-end latency are given for $\mathcal{M}_{\text{global}}$ in the following theorem.

**Theorem 2.** *For the tandem network in Fig. 1 with n relay nodes and $m_h$ channels in the $h^{\text{th}}$ hop, the minimum end-to-end latency with $\mathcal{M}_{\text{global}}$ is*

$$\tau_n^* = \frac{\left(C_{n,m_n}^{-1} + \tau_{n-1}^*\right) \prod_{k=1}^{m_n - 1} C_{n,k+1} \left(C_{n,k}^{-1} + \tau_{n-1}^*\right)}{1 + \sum_{i=2}^{m_n} \prod_{k=1}^{i-1} C_{n,k+1} \left(C_{n,k}^{-1} + \tau_{n-1}^*\right)}, \quad (22)$$

*with initial condition $\tau_0^* \triangleq \left(\sum_{i=1}^{m_0} C_{0,i}\right)^{-1}$, achieved by*

$$\alpha_{h,k} = \begin{cases} C_{h,k} \left(\sum_{k=1}^{m_h} C_{h,k}\right)^{-1}, & h = 0 \\ \frac{\prod_{i=1}^{k-1} C_{h,i+1} \left(C_{h,i}^{-1} + \tau_{h-1}^*\right)}{1 + \sum_{j=2}^{m_h} \prod_{i=1}^{j-1} C_{h,i+1} \left(C_{h,i}^{-1} + \tau_{h-1}^*\right)}, & h \geq 1. \end{cases} \quad (23)$$

*Proof.* For $h \in \{0\} \cup [n]$, we denote by $\tau_h^*$ the minimum end-to-end latency from hop 0 to hop $h$. In addition, we define the effective capacity as the reciprocal of minimum end-to-end latency, i.e., $\mathcal{E}_h \triangleq \left(\tau_h^*\right)^{-1}$. According to Lemma 1, the initial effective capacity $\mathcal{E}_0$ is given as $\mathcal{E}_0 = \sum_{i=1}^{m_0} C_{0,i}$.

For the $h^{\text{th}}$ hop with $h \geq 1$, we lump hops 0 to $h - 1$ together, with effective capacity $\mathcal{E}_{h-1}$. By Lemma 2, we know that at the relay node the local effective capacity depends on the its connected channels on two hops. More precisely, if the resulting effective capacity by lumping hops 0 to $h - 1$ is updated to $C_0$ in Fig. 3, then the effective capacity at the $(h + 1)^{\text{th}}$ node can be obtained by treating all channels in the $h^{\text{th}}$ hop equal to those in the source-relay hop in Fig. 3. Thus, the effective capacity for the concatenated system with $h$ hops is expressed as

$$\mathcal{E}_h = \frac{1 + \sum_{i=2}^{m_h} \prod_{k=1}^{i-1} C_{h,k+1} \left(C_{h,k}^{-1} + \mathcal{E}_{h-1}^{-1}\right)}{\left(C_{h,m_h}^{-1} + \mathcal{E}_{h-1}^{-1}\right) \prod_{k=1}^{m_h - 1} C_{h,k+1} \left(C_{h,k}^{-1} + \mathcal{E}_{h-1}^{-1}\right)}. \quad (24)$$

With the recursive expression of $\mathcal{E}_h$, we can finally obtain $\tau_n^* = \mathcal{E}_n^{-1}$ when $h$ reaches $n$. $\square$

We know from Theorem 2 that in the middle phase of the transmission the file is distributed in all buffer-aided relay nodes rather than stacked in one buffer, and all relay nodes can simultaneously forward the buffered fractions to the subsequent nodes. The advantage of applying $\mathcal{M}_{\text{global}}$ is that a longer queue at a single relay node (i.e., a bottleneck) can be avoided. Meanwhile, concurrent transmissions at all relay nodes enable even utilizations among all buffers, thereby resulting in higher transmission efficiency.

In Theorem 2, we find that the traffic allocation with $\mathcal{M}_{\text{global}}$ is found through a recursion. In contrast to the exhaustive method for searching the optimal solution, this recursive scheme significantly decreases the computational complexity. Intuitively, the traffic allocation at each node in the network is a function of the delay in the sub-network consisting of the subsequent nodes, while disregarding exact traffic allocations over the following hops. Hence, the resulting latency from the present node to the end can be treated as a whole and adopted for computing the latency from its previous node, thereby indicating the recursive property when applying $\mathcal{M}_{\text{global}}$.

*C. Discussion of Computational Complexity*

Note that the analysis in this paper is conducted for a given channel capacity. Assuming that the cost for acquiring the information of each channel, i.e., channel estimation, signaling, or beamforming, is fixed (or upper bounded), then the overall cost can be characterized by the number of channels in total for performing the optimized traffic allocation, i.e., the computational complexity.

In what follows, we determine the computational complexity for $\mathcal{M}_{\text{local}}$ and $\mathcal{M}_{\text{global}}$. With respect to the multi-hop network in Fig. 1, we know that:

- Using $\mathcal{M}_{\text{local}}$, there are $m_h$ channels considered for computing the traffic allocation at node $h+1$ for any $h \in \{0\} \cup [n]$ within each hop. Therefore, for the whole network, the overall computational complexity is in $O\left(\sum_{h=0}^{n} m_h\right)$.
- Using $\mathcal{M}_{\text{global}}$, within each hop there are $\sum_{i=0}^{h} m_i$ channels considered for computing the traffic allocation with $\mathcal{M}_{\text{global}}$ at node $h+1$ for any $h \in \{0\} \cup [n]$. Thus, for the whole network, the overall computational complexity is in $O\left(\sum_{h=0}^{n} \sum_{i=0}^{h} m_i\right)$.

Due to the fact that $\sum_{h=0}^{n} m_h \leq \sum_{h=0}^{n} \sum_{i=0}^{h} m_i$ (the equality holds only if $n = 0$), we have $O\left(\sum_{h=0}^{n} m_h\right) \subset O\left(\sum_{h=0}^{n} \sum_{i=0}^{h} m_i\right)$, which indicates that the overall computational complexity for $\mathcal{M}_{\text{local}}$ is lower. The comparison of the overall complexity to perform two traffic allocation schemes will be elaborated on in the next section.

## IV. Performance Gain by Global Allocation

In this section, we study the performance gain by adopting $\mathcal{M}_{\text{global}}$, relative to that by adopting $\mathcal{M}_{\text{local}}$. For the sake of fairness and convenience, we here restrict ourselves to a homogeneous version of the multi-hop networks in Fig. 1. That is, for all $h \in \{0\} \cup [n]$ and $k \in [m]$, each hop has the same number of channels, i.e., $m_h = m$, and the capacity of all channels are identical, i.e., $C_{h,k} = C$.

*A. Relative Performance Gain and Its Asymptote*

Given $n$ relay nodes and $m$ channels per hop, taking $\mathcal{M}_{\text{global}}$ as the reference, we define the relative performance gain by $\mathcal{M}_{\text{global}}$ as

$$\rho(n,m) \triangleq \frac{\tau_n^*|_{\text{local}}}{\tau_n^*|_{\text{global}}}, \quad (25)$$

where $\tau_n^*|_{\text{global}}$ and $\tau_n^*|_{\text{local}}$ represent the minimum end-to-end latencies obtained in Theorem 2 and Theorem 1, respectively.

We next present a result on the relative performance gain $\rho(n,m)$.

**Theorem 3.** *Given $n$ relay nodes and $m$ channels in each hop, the relative performance gain $\rho(n,m)$ is*

$$\rho(n,m) = \frac{n+1}{m} \left(u_n^*\right)^{-1}, \quad (26)$$

*where $u_k^*$ for all $k \in [n]$ is given as*

$$u_k^* = \left(1 - \left(1 + u_{k-1}^*\right)^{-m}\right)^{-1} u_{k-1}^* \quad (27)$$

*with initial condition $u_0^* = m^{-1}$.*

*Proof.* Please see Appendix C. □

Based on Theorem 3, it is interesting to study the relative performance gain when the number of channels per hop increases. With any given number of relay nodes $n$, the asymptotic relative performance gain $\bar{\rho}(n)$ is defined as

$$\bar{\rho}(n) \triangleq \lim_{m \to \infty} \rho(n,m). \quad (28)$$

The following Theorem 4 gives an expression for $\bar{\rho}(n)$.

**Theorem 4.** *Given $n$ relay nodes, the asymptotic performance gain $\bar{\rho}(n)$ for any $n \geq 0$ is recursively given as*

$$\bar{\rho}(n) = \frac{n+1}{n} \left(1 - \exp\left(-\frac{n}{\bar{\rho}(n-1)}\right)\right) \bar{\rho}(n-1), \quad (29)$$

*with initial condition $\bar{\rho}(0) = 1$.*

*Proof.* Please see Appendix D. □

Theorem 4 demonstrates that the relative performance gain approaches a constant that only depends on the number of relay nodes, when the number of channels in each hop goes to infinity. This limiting performance quantifies the maximum relative performance gain, achieved by letting the number of channels per hop grow to infinity.

*B. Gain-Complexity Trade-off*

Considering the overall computational complexity for the different traffic allocation schemes, for the homogeneous setting in this section, it is easy to obtain that the number of channels for traffic allocation with $\mathcal{M}_{\text{local}}$, denoted as $f_{\text{local}}(m,n)$, is given by

$$f_{\text{local}}(m,n) = m(n+1), \quad (30)$$

while that with $\mathcal{M}_{\text{global}}$, denoted as $f_{\text{global}}(m,n)$, is given by

$$f_{\text{global}}(m,n) = \frac{m(n+1)(n+2)}{2} = \frac{n+2}{2} \cdot f_{\text{local}}(m,n). \quad (31)$$

Thus, evidently, the relative overall computational complexity, which can be defined via comparing $\mathcal{M}_{\text{global}}$ to $\mathcal{M}_{\text{global}}$, grows with the number of relay nodes, linearly, i.e.,

$$\frac{f_{\text{global}}(m,n)}{f_{\text{local}}(m,n)} = \frac{n+2}{2} \in O(n), \quad (32)$$

which only depends on the number of relay nodes. Jointly with Theorem 4, we can see that $\mathcal{M}_{\text{global}}$ achieves a relative performance gain of $\bar{\rho}(n)$ at the expense of an $n$ times higher computational complexity. Thus, there exists a trade-off between the relative performance gain and the overall computational complexity by global allocation, i.e., between $\bar{\rho}(n)$ and $O(n)$.

## V. AVERAGE LATENCY FOR TWO-HOP LINEAR MM-WAVE NETWORKS WITH NAKAGAMI-$m$ FADING

In this section, we focus on the average end-to-end latency for a two-hop mm-wave network as shown in Fig. 3, with two independently fading channels between the source and the relay node. We consider small-scale fading for all channels in the two-hop network and assume independent block fading. That is, for each fraction and each hop, the channel is independent and identically distributed (i.i.d.) but constant during the delivery of the file fraction over each corresponding channel. We assume that the instantaneous channel state information (CSI) is known at the node that makes the resource allocation. Hence, the traffic allocation is performed after acquiring the CSI[3].

Observing the form of the end-to-end latency in Theorem 1 or Theorem 2, we notice that it is rather difficult to derive a closed-form expression of the average latency, since the end-to-end latency is a reciprocal of the end-to-end effective capacity. For the sake of tractability, we aim at lower bounds to characterize the average latency performance. In what follows, we first give the average capacity for mm-wave channels with Nakagami-$m$ fading, and we subsequently derive the lower bounds on the end-to-end latency when using $\mathcal{M}_{\text{local}}$ and $\mathcal{M}_{\text{global}}$, respectively.

### A. Average Channel Capacity

It has been reported in [6] that, unlike the channel characteristics in sub-6 GHz bands, the small-scale fading in mm-wave channels is not significant, due to the adoption of highly directional antennas and the weak capability of reflection/diffraction. For tractability, in this paper, we assume that the amplitude of mm-wave channel coefficient follows Nakagami-$m$ fading, as in [30]. Hence, for a given signal-to-noise (SNR) $\xi$, the normalized capacity of a mm-wave channel with Nakagami-$m$ fading can be written as

$$C = \log_2(1 + g \cdot \xi), \tag{33}$$

where the random variable $g$ represents the channel power gain, which follows the gamma distribution, i.e., $g \sim \Gamma(M, M^{-1})$ with a positive Nakagami parameter $M$. The variance is $M^{-1}$, hence the randomness decreases with $M$, and the channel becomes deterministic as $M \to \infty$. $M = 1$ corresponds to Rayleigh fading.

We assume that mm-wave channels $C_i$, $i \in \{0, 1, 2\}$, have the identical Nakagami parameter[4] $M$. Then, with the aid of Meijer G-function, the average capacity for channels with Nakagami-$m$ fading can be obtained as

$$\mathbb{E}[C_i] = \int_0^\infty \log_2(1 + x\xi_i) f(x; M) dx$$
$$= \frac{M^M}{\xi_i^M \Gamma(M) \ln(2)} \cdot G_{2,3}^{3,1}\left(\begin{matrix} -M, 1-M \\ 0, -M, -M \end{matrix} \middle| \frac{M}{\xi_i}\right), \tag{34}$$

---

[3]In this paper we only consider the case of perfect CSI to study the best possible performance. If only the statistical CSI or no CSI is available, the performance by using $\mathcal{M}_{\text{local}}$ and $\mathcal{M}_{\text{global}}$ is inevitably degraded.

[4]Normally, the randomness in mm-wave channels is relatively weak, such that the Nakagami parameter $M \geq 3$ as in [30], [31].

where $\mathbb{E}[\cdot]$ denotes the expectation operator, and $\xi_i$ is the SNR on $C_i$. Here, $G_{p,q}^{m,n}\left(\begin{matrix} a_1, a_2, ..., a_p \\ b_1, b_2, ..., b_q \end{matrix} \middle| z\right)$ denotes the Meijer G-function [32], where $0 \leq m \leq q$ and $0 \leq n \leq p$, and parameters $a_j$, $b_j$ and $z \in \mathbb{C}$.

### B. Lower Bounds on Average Latency

In this subsection, for different traffic allocation schemes, we derive lower bounds on the average end-to-end latency, in terms of $\mathbb{E}[C_i]$ for $i \in \{0, 1, 2\}$ (as in the previous subsection).

*1) Using $\mathcal{M}_{\text{local}}$:* Given $C_0$, $C_1$ and $C_2$, applying the traffic allocation by Theorem 1 to (2), it is easy to obtain that the minimum end-to-end latency with $\mathcal{M}_{\text{local}}$ is

$$\tau_2^* = C_0^{-1} + (C_1 + C_2)^{-1}. \tag{35}$$

Then, a lower bound on the expectation of $\tau_2^*$ is presented in the following proposition.

**Proposition 1.** *A lower bound on the minimum average end-to-end latency for $\mathcal{M}_{\text{local}}$ is*

$$\mathbb{E}[\tau_2^*] \geq (\mathbb{E}[C_0])^{-1} + (\mathbb{E}[C_1 + C_2])^{-1}. \tag{36}$$

*Proof.* Based on $\tau_2^*$ given in (35), the minimum average end-to-end latency can be expressed as $\mathbb{E}[\tau_2^*] = \mathbb{E}[C_0^{-1}] + \mathbb{E}[(C_1 + C_2)^{-1}]$. Regarding the existence of $\mathbb{E}[C_0^{-1}]$ for $M > 1$, we note that

$$\mathbb{E}[C_0^{-1}] \leq \ln(2)\left(\frac{1}{2} + \xi_0^{-1}\mathbb{E}[g^{-1}]\right)$$
$$= \ln(2)\left(\frac{1}{2} + \xi_0^{-1}\left(1 + (M-1)^{-1}\right)\right) < \infty, \tag{37}$$

where the first line applies the following inequality for $x \geq 0$: $\ln(1+x) \geq \frac{2x}{2+x}$. Therefore, the existence $\mathbb{E}[C_0^{-1}]$ is guaranteed.

Finally, by applying Jensen's inequality, i.e., $\mathbb{E}[X] \cdot \mathbb{E}[X^{-1}] \geq 1$ for positive random variables $X$, we can obtain that $\mathbb{E}[\tau_2^*] \geq (\mathbb{E}[C_0])^{-1} + (\mathbb{E}[C_1 + C_2])^{-1}$, which completes the proof. □

*2) Using $\mathcal{M}_{\text{global}}$:* given $C_0$, $C_1$ and $C_2$, applying the traffic allocation by Theorem 2 to (2), the minimum end-to-end latency with $\mathcal{M}_{\text{global}}$ is

$$\tau_2^* = \frac{(C_0^{-1} + C_1^{-1})(C_0^{-1} + C_2^{-1})}{C_0^{-1} + C_1^{-1} + C_2^{-1}}. \tag{38}$$

Then, a lower bound on the expectation of $\tau_2^*$ is given in the following proposition.

**Proposition 2.** *A lower bound on the minimum end-to-end latency for $\mathcal{M}_{\text{global}}$ is*

$$\mathbb{E}[\tau_2^*] \geq (\mathbb{E}[C_0])^{-1} + (\mathbb{E}[C_1 + C_2] + \epsilon_0 \mathbb{E}[C_1 C_2])^{-1}, \tag{39}$$

*where $\epsilon_0$ for $M > 1$ is defined as*

$$\epsilon_0 \triangleq \ln(2)\left(\frac{1}{2} + \xi_0^{-1}\left(1 + (M-1)^{-1}\right)\right). \tag{40}$$

*Proof.* With $\tau_2^*$ given in (38), we know that

$$\mathbb{E}[\tau_2^*] = \mathbb{E}[C_0^{-1}] + \mathbb{E}\left[\left(C_1 + C_2 + C_0^{-1}C_1C_2\right)^{-1}\right]$$
$$\geq (\mathbb{E}[C_0])^{-1} + \mathbb{E}^{-1}[C_1 + C_2 + C_0^{-1}C_1C_2], \tag{41}$$

where the second line is achieved by using Jensen's inequality. We notice that $\mathbb{E}\left[C_0^{-1}\right]$ is upper bounded as $\mathbb{E}\left[C_0^{-1}\right] \leq \ln(2)\left(\frac{1}{2} + \xi_0^{-1}\left(1 + (M-1)^{-1}\right)\right) \triangleq \epsilon_0$. □

For $M = 1$, we notice that $\mathbb{E}\left[C_0^{-1}\right]$ does not exist since $\int_0^\infty \left(\log_2(1 + \xi x)\right)^{-1} \exp(-x)\,dx$ does not converge. Hence, the bounding techniques in Proposition 1 and Proposition 2 are not applicable to the scenarios with Rayleigh fading channels ($M = 1$). Fortunately, this case is less relevant for mm-wave communications.

It is worth mentioning that the method for analyses above can be extended to mm-wave networks with more relay nodes and more channels in each hop: The lower bound $\mathcal{M}_{\text{local}}$ can be obtained by performing Jensen's inequality on the component latency in each hop, while the lower bound with respect to $\mathcal{M}_{\text{global}}$ can be obtained by following the recursive expression for end-to-end latency presented in Theorem 2.

## VI. Performance Evaluation

In this section, we will evaluate the end-to-end latency for $\mathcal{M}_{\text{local}}$ and $\mathcal{M}_{\text{global}}$. The performance evaluation consists of the following two parts:

(i) We focus on the allocation schemes $\mathcal{M}_{\text{local}}$ and $\mathcal{M}_{\text{global}}$ and investigate their corresponding performance. With simulations, we first validate allocation schemes developed in Theorem 1 and Theorem 2 for the minimum end-to-end latency. After validating the allocation schemes, we subsequently provide numerical results focusing on Theorem 3 and Theorem 4 and assess the performance achieved by two distinct schemes.

(ii) Following the two-hop network adopted in Sec. V, we evaluate the average end-to-end latency performance in the presence of Nakagami-$m$ fading in mm-wave channels. We first show the tightness of the lower bounds derived in Proposition 1 and Proposition 2. Further discussions related to the average performance are presented afterwards.

We assume that the size of the transmitted file is normalized to 1 without loss of generality. Other system settings for the above two assessments will be elaborated on.

### A. Performance of the Two Allocation Schemes

Focusing on the performance of two traffic allocation schemes, we assume deterministic channels, such that the channel capacity is treated as constants. Furthermore, for fairness and simplicity, we follow the homogeneous setting used in Sec. IV, i.e., $m_h = m$ and $C_{h,k} = C$ for all $h \in \{0\} \cup [n]$ and $k \in [m]$.

We simulate the end-to-end latency of a two-hop system with two channels per hop, i.e., $n = 1$ and $m = 2$, and the performance for $\mathcal{M}_{\text{local}}$ and $\mathcal{M}_{\text{global}}$ is shown in Fig. 4. For traffic allocations $\underline{\alpha}_0$ (at the relay node) and $\underline{\alpha}_1$ (at the source), we consider variables $\alpha_{0,1} \in [0, 0.5]$ and $\alpha_{1,1} \in [0, 0.5]$, and the remaining allocations can be characterized in terms of $\alpha_{0,1}$ and $\alpha_{1,1}$, i.e., $\alpha_{0,2} = 1 - \alpha_{0,1}$ and $\alpha_{1,2} = 1 - \alpha_{1,1}$, respectively, due to the fact $m = 2$. In both Fig. 4(a) and Fig. 4(b), we vary $\alpha_{0,1}$ and $\alpha_{1,1}$, jointly. In general, it is evident that the resulting latencies by distinct allocation schemes are different. We can see in Fig. 4(a) that the minimum end-to-end latency is 1 when applying $\mathcal{M}_{\text{local}}$, which is achieved at $\alpha_{0,1} = 0.5$ and $\alpha_{1,1} = 0.5$. However, in Fig. 4(b), the minimum end-to-end latency is 0.9 when applying $\mathcal{M}_{\text{global}}$, which is achieved at $\alpha_{0,1} = 0.5$ and $\alpha_{1,1} = 0.4$. The optimal traffic allocations enabling the minimum end-to-end latency observed from Fig. 4 are in accordance with our analytical results derived in Theorem 1 and Theorem 2.

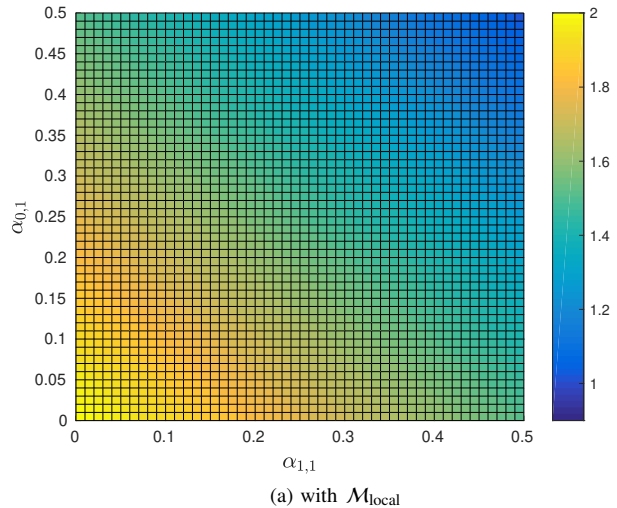

(a) with $\mathcal{M}_{\text{local}}$

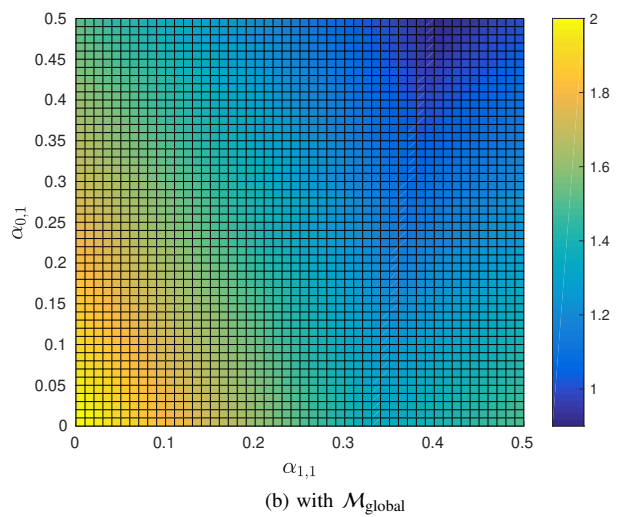

(b) with $\mathcal{M}_{\text{global}}$

Fig. 4. End-to-end latency performance with $\mathcal{M}_{\text{local}}$ and $\mathcal{M}_{\text{global}}$, where the number of relay nodes is $n = 1$, the number of channels per hop is $m = 2$, the capacity of each channel is $C = 1$, and traffic allocation $\alpha_{0,1}$ and $\alpha_{1,1}$ both vary from 0 to 0.5.

In Fig. 5, we investigate the minimum end-to-end latency $\tau^*$ against the number of relay nodes $n$, where different numbers of per-hop channels $m$ are considered. For both $\mathcal{M}_{\text{local}}$ and $\mathcal{M}_{\text{global}}$, we find that $\tau^*$ can be significantly reduced when elevating $m = 2$ to $m = 10$. This coincides with the intuition that increasing the number of channels is equivalent to producing a larger effective channel capacity in each hop, which in turn leads to a lower latency for file delivery. Besides, we notice that the benefit by adopting $\mathcal{M}_{\text{global}}$ becomes remarkable as $n$ increases. For instance, for $m = 2$, compared to $\mathcal{M}_{\text{local}}$, applying $\mathcal{M}_{\text{global}}$ reduces the latency by 25% at $n = 3$, while the reduction is enlarged to 40% at $n = 9$. The performance improvement shown above stems from the efficient utilization of buffers at relay nodes in $\mathcal{M}_{\text{global}}$, since long queues are avoided by performing the optimal traffic allocations globally at all relay nodes. This observation reveals the great advantage of $\mathcal{M}_{\text{global}}$ in networks with more relay nodes. Furthermore, we can see that there is an intersection at $n = 6$, between the curve with $m = 5$ for $\mathcal{M}_{\text{local}}$ and the curve with $m = 10$ for $\mathcal{M}_{\text{global}}$. This finding indicates that $\mathcal{M}_{\text{global}}$ with fewer channels is still competitive in outperforming $\mathcal{M}_{\text{local}}$ with more channels as long as there are sufficiently many relay nodes, which again highlights the benefit of global allocation.

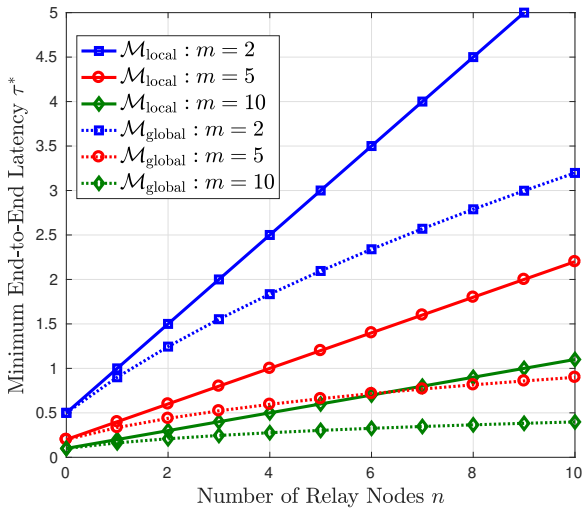

Fig. 5. Minimum end-to-end latency $\tau^*$ vs. number of relay nodes $n$, where the number of channels per hop is $m = 2, 5$ or $10$, the capacity of each channel is $C = 1$, and the size of transmitted file is 1.

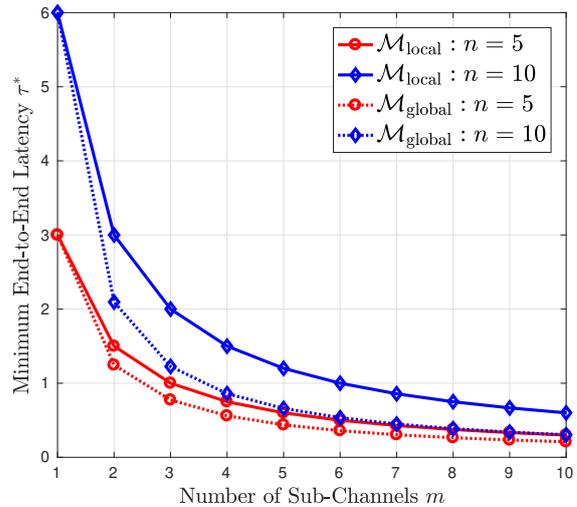

Fig. 6. Minimum end-to-end latency $\tau^*$ vs. number of channels $m$, where the number of relay nodes is $n = 5$ or $10$, the capacity of each channel is $C = 1$, and the size of transmitted file is 1.

Given $n$ relay nodes, the minimum end-to-end latency $\tau^*$ against the number of channels $m$ is illustrated in Fig. 6. For both $\mathcal{M}_{\text{local}}$ and $\mathcal{M}_{\text{global}}$, $\tau^*$ is dramatically reduced at the beginning of increasing $m$, while the decaying rates slow down when $m$ becomes large, i.e., when $m \geq 5$. This finding indicates that it is definitely beneficial to have multiple channels for reducing the end-to-end latency, but the benefit diminishes as the number of channels increases. Therefore, in practice, considering the cost of system implementations, it is not necessary to increase the number of channels above about 8.

In Fig. 7, we investigate the relative performance gain $\rho(n, m)$ with respect to the number of relay nodes $n$, where the number of per-hop channels $m$ varies from $m = 1$ to $\infty$. We find that there is no difference between $\mathcal{M}_{\text{local}}$ and $\mathcal{M}_{\text{global}}$ when $m = 1$, i.e., $\rho(n, 1) = 1$ for all $n$, since neither the local allocation nor the global allocation is actually performed if there is only one single channel per hop. However, $\rho(n, m)$ obviously increases when $m$ or $n$ grows. This indicates the substantial advantage of $\mathcal{M}_{\text{global}}$ compared to $\mathcal{M}_{\text{local}}$, especially when the number of channels per hop or the number of relay nodes is large. In addition, when $m \to \infty$, asymptotic performance gain $\bar{\rho}(n)$ characterizes the upper bound of the relative benefits. For instance, at $n = 20$, the latency can be reduced to 20% of $\mathcal{M}_{\text{local}}$ at the most, when applying $\mathcal{M}_{\text{global}}$. The asymptote in Fig. 7 is obtained from Theorem 4.

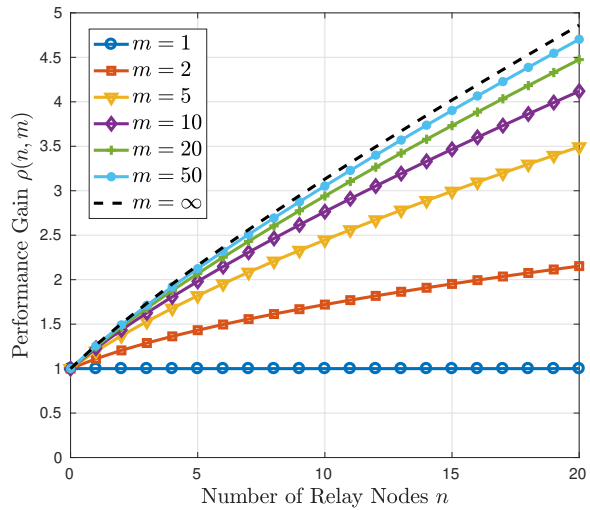

Fig. 7. Relative performance gain $\rho(n, m)$ vs. number of relay nodes $n$, where the number of channels per hop is $m = 1, 2, 5, 10, 20, 50$ or $\infty$.

### B. Average Latency in Millimeter-wave Networks

To investigate the average end-to-end latency of the two-hop mm-wave network in Sec. V (see Fig. 3), we consider a network where the relay node is deployed on the line between the source and the destination (All networks considered here are linear in the sense of the network topology.). Starting from the source, we denote by $r$ and $L$ the (normalized) distances to the relay node and to the destination, respectively. Besides, we assume that the transmit power at the source and the relay node are both $\gamma$, and the power of the background noise is set to 1 without loss of generality. Applying the path loss model for line-of-sight (LOS) mm-wave communications [6], [8], the SNR $\xi_i$ in $C_i = \log_2\left(1 + g_i \xi_i\right)$ can be written as

$$\xi_i = \begin{cases} \gamma (L - r)^{-\alpha}, & i = 0 \\ \gamma r^{-\alpha}, & i \in \{1, 2\}, \end{cases} \quad (42)$$

where $\alpha$ denotes the path loss exponent. The gamma-distributed random variables $g_i$ for $i \in \{0, 1, 2\}$ are independent and identically distributed with Nakagami parameter $M$.

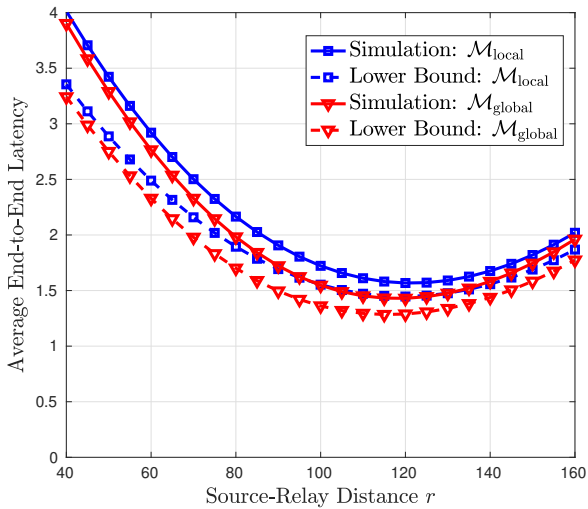

Fig. 8. Average end-to-end latency $\mathbb{E}\left[\tau_2^*\right]$ and the lower bounds, against source-relay distance $r$, where (normalized) source-destination distance $L = 200$, transmit power $\gamma = 60$ dB, path loss exponent $\alpha = 3$, and Nakagami parameter $M = 5$.

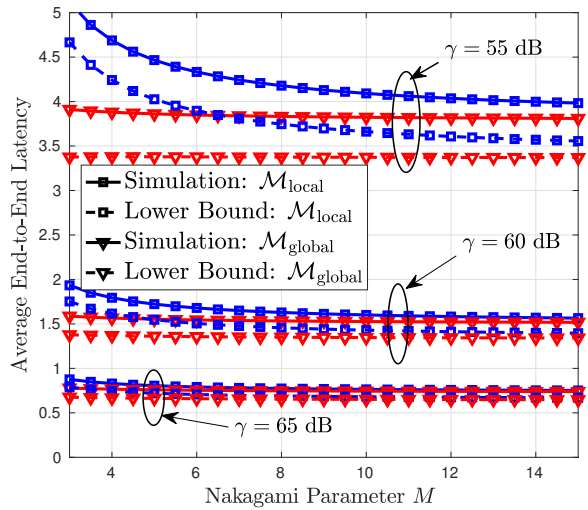

Fig. 9. Average end-to-end latency $\mathbb{E}\left[\tau_2^*\right]$ and the lower bounds against Nakagami parameter $M$, where (normalized) source-destination distance $L = 200$, (normalized) source-relay distance $r = 100$, transmit power $\gamma = 55$ dB, 60 dB or 65 dB, and path loss exponent $\alpha = 3$.

With Nakagami fading in mm-wave channels, the average end-to-end latency and the lower bounds are illustrated in Fig. 8. From the simulation results for the two traffic allocation schemes, it can be seen that the lower bounds given in Proposition 1 and Proposition 2 are quite tight. Furthermore, the average latency first decreases to the minimum and subsequently increases, when $r$ grows from 40 to 160. This observation indicates the critical role of relay deployment in minimizing the end-to-end latency. We can also see that the minimum average latency for $\mathcal{M}_{\text{local}}$ is obtained roughly at $r = 120$, while the minimum average latency for $\mathcal{M}_{\text{global}}$ is obtained roughly at $r = 115$. This slight difference tells that different relay deployments may be needed for different allocation schemes.

In Fig. 9 we show the average end-to-end latency $\mathbb{E}\left[\tau_2^*\right]$ against varying Nakagami parameter $M$. As aforementioned, a larger $M$ corresponds to a more deterministic channel. We see from Fig. 9 that for both $\mathcal{M}_{\text{local}}$ and $\mathcal{M}_{\text{global}}$ the average end-to-end latency $\mathbb{E}\left[\tau_2^*\right]$ decreases as $M$ grows from 3 to 15, while the reduction in latency by increasing $M$ gradually diminishes. Moreover, the simulation results gradually approach the corresponding lower bounds when $M$ increases. This is due to the fact that Jensen's inequality in Proposition 1 or Proposition 2 gives a tighter lower bound $(\mathbb{E}\left[X\right])^{-1}$ for $\mathbb{E}\left[X^{-1}\right]$ when mm-wave channels become more deterministic (higher $M$). In addition, we find that the tightness of the lower bounds for both allocation schemes are improved when $\gamma$ increases from 55 dB to 65 dB. Thus, the lower bounds get even tighter when the transmit powers are high.

## VII. Conclusions

We have studied the end-to-end latency in multi-hop mm-wave networks by applying two traffic allocation schemes, namely local allocation and global allocation. In our networks, buffers are equipped at the source node and the relay nodes, and multiple independent channels exist in each hop. For given channel capacities, we have provided closed-form expressions of the end-to-end latency for the two allocation schemes and quantified the advantages of the global allocation scheme relative to the local one. Some asymptotic analyses have also been performed. Compared to local allocation, the advantage of global allocation grows as the number of relay nodes $n$ increases, at the expense of an $n$ times higher computational complexity. Besides, increasing the number of channels monotonically decreases the latency, which asymptotically reaches a constant that depends only on the number of relay nodes. Furthermore, taking a specific two-hop linear mm-wave network as an example, we have derived tight lower bounds on the average end-to-end latency for two traffic allocation schemes with Nakagami-$m$ fading incorporated. We have also noticed the great importance of proper deployment of the relay node. These results can provide insights for designing or implementing low-latency multi-hop mm-wave networks.

# APPENDIX A
## PROOF OF LEMMA 2

Due to the buffer at the relay node, fractions from distinct channels are first stacked in the queue and subsequently pushed on the channel connecting the destination. To simplify the notation, we assume that fraction $\alpha_j$ arrives at the buffer-aided relay node prior to fraction $\alpha_k$ if $j \leq k$, for all $j, k \in [m]$, without loss of generality. Letting $w_i$ for all $i \in [m]$ denote the latency of fraction $\alpha_i$ traversing from the source to the destination, according to the arrival orders at the buffer-aided relay node, we have $w_1 = \alpha_1 \left( C_1^{-1} + C_0^{-1} \right)$ and $w_2 = \max \{w_1, \alpha_2 C_2^{-1}\} + \alpha_2 C_0^{-1}$. In light of above, we can express the component delay $w_i$ for $i \in [m]$ in general as $w_i = \max \{w_{i-1}, \alpha_i C_i^{-1}\} + \alpha_i C_0^{-1}$ with the initial condition $w_0 = 0$.

It is evident that $w_j < w_k$ for any $j < k$, since $w_k \geq w_{k-1} + \alpha_k C_0^{-1} > w_{k-1} > \ldots > w_j$. Thus, latency $\tau$ is reduced to $w_m$, i.e., $\max_{1 \leq i \leq m} \{w_i\} = w_m$, and the minimum latency $\tau^*$ can be expressed as $\tau^* = \min_{\|\alpha\|_1 = 1} w_m$. Applying the recursion for $w_m$, we equivalently have

$$\begin{aligned}
\tau^* &= \min_{\|\alpha\|_1 = 1} \max \{w_{m-1}, \alpha_m C_m^{-1}\} + \alpha_m C_0^{-1} \\
&= \min_{\alpha_m \in (0,1)} \min_{\sum_{i=1}^{m-1} \alpha_i = 1 - \alpha_m} \max \{w_{m-1}, \alpha_m C_m^{-1}\} + \alpha_m C_0^{-1} \\
&= \min_{\alpha_m \in (0,1)} \max \left\{ \min_{\sum_{i=1}^{m-1} \alpha_i = 1-\alpha_m} \{w_{m-1}\}, \alpha_m C_m^{-1} \right\} + \alpha_m C_0^{-1},
\end{aligned} \quad (43)$$

where the last line is obtained based on the fact that $w_{m-1}$ depends on $\{\alpha_i\}$ for all $i \in [m-1]$, while $\alpha_m C_0^{-1}$ and $\alpha_m C_m^{-1}$ can be treated as constants with respect to a given $\alpha_m$.

For notational simplicity, we define

$$\xi^* \triangleq \min_{\sum_{i=1}^{m-1} \alpha_i = 1} w_{m-1}, \quad (44)$$

which denotes the optimized latency of delivering one normalized-size file in the network with $m-1$ channels between the source and the relay node. Thanks to the linear mapping between the allocated traffic loads and the resulting latency, with respect to any $z > 0$, we can easily obtain that

$$\min_{\sum_{i=1}^{m-1} \alpha_i = z} w_{m-1} = \min_{\sum_{i=1}^{m-1} \alpha_i = 1 \cdot z} w_{m-1} = z \cdot \xi^*. \quad (45)$$

Therefore, $\tau^*$ can be further reduced to

$$\begin{aligned}
\tau^* &= \min_{\alpha_m \in (0,1)} \max \{(1-\alpha_m)\xi^*, \alpha_m C_m^{-1}\} + \alpha_m C_0^{-1} \\
&= \min_{\alpha_m \in (0,1)} \max \{\xi^* - \alpha_m(\xi^* - C_0^{-1}), \alpha_m(C_m^{-1} + C_0^{-1})\}.
\end{aligned} \quad (46)$$

Note that $\xi^*$ denotes the latency for delivering one normalized-size file from the source to the destination via the relay node, while $C_0^{-1}$ denotes the latency for delivering the file from the relay node to the destination. The former is strictly greater than the latter, i.e., $\xi^* > C_0^{-1}$. Then, we can see that $\xi^* - \alpha_m(\xi^* - C_0^{-1})$ is monotonically increasing with $\alpha_m$, while $\alpha_m(C_m^{-1} + C_0^{-1})$ is monotonically decreasing with $\alpha_m$. Hence, $\tau^*$ is obtained whenever $\xi^* - \alpha_m(\xi^* - C_0^{-1}) = \alpha_m(C_m^{-1} + C_0^{-1})$.

One particular solution that meets the condition shown above for minimizing the latency is to $w_m = \alpha_m C_m^{-1}$. Iteratively, $w_{i-1} = \alpha_i C_i^{-1}$ should hold for all $i \in [m]$. In this case, we can obtain that $\alpha_2 C_2^{-1} = \alpha_1 \left( C_1^{-1} + C_0^{-1} \right)$, and the general expression for $2 \leq i \leq m$ is

$$\alpha_i = \alpha_1 \prod_{k=1}^{i-1} C_{k+1} \left( C_k^{-1} + C_0^{-1} \right). \quad (47)$$

Paired with the constraint $\|\alpha\|_1 = 1$, we can immediately solve $\alpha_1$ as

$$\alpha_1 = \left( 1 + \sum_{i=2}^{m} \prod_{k=1}^{i-1} C_{k+1} \left( C_k^{-1} + C_0^{-1} \right) \right)^{-1}. \quad (48)$$

Thus, applying the recursive expression of $\alpha_i$, we have

$$\alpha_m = \frac{\prod_{k=1}^{m-1} C_{k+1} \left( C_k^{-1} + C_0^{-1} \right)}{1 + \sum_{i=2}^{m} \prod_{k=1}^{i-1} C_{k+1} \left( C_k^{-1} + C_0^{-1} \right)}, \quad (49)$$

which further gives the minimum latency $\tau^*$ as

$$\tau^* = \frac{\left( C_m^{-1} + C_0^{-1} \right) \prod_{k=1}^{m-1} C_{k+1} \left( C_k^{-1} + C_0^{-1} \right)}{1 + \sum_{i=2}^{m} \prod_{k=1}^{i-1} C_{k+1} \left( C_k^{-1} + C_0^{-1} \right)}. \quad (50)$$

# APPENDIX B
## PROOF OF COROLLARY 1

We define the multinominal $C(i, m)$ as

$$C(i, m) \triangleq \sum_{k_1, \ldots, k_i \in [m]}^{\neq} \prod C_{k_j}, \quad (51)$$

where the $\neq$ indicates that $k_u$ and $k_v$ are not equal for any $u, v \in [m]$. Rewriting the expression of $\tau^*$ in Lemma 2, we can obtain that

$$\begin{aligned}
\tau^* &= \frac{\sum_{i=0}^{m} C_0^{m-i} \cdot C(i,m)}{C_0 \sum_{i=1}^{m} C_0^{m-i} \cdot C(i,m)} = \frac{C_0^m + \sum_{i=1}^{m} C_0^{m-i} \cdot C(i,m)}{C_0 \sum_{i=1}^{m} C_0^{m-i} \cdot C(i,m)} \\
&= C_0^{-1} + \left( \sum_{i=1}^{m} C_0^{1-i} C(i,m) \right)^{-1}.
\end{aligned} \quad (52)$$

Subsequently, the minimization of $\tau^*$ with respect to the constraint $\|\underline{C}\|_1 \triangleq \sum_{i=1}^{m} C_i = 1$ can be reformulated as

$$\min_{\|\underline{C}\|_1 = 1} \tau^* = C_0^{-1} + \left( \sum_{i=1}^{m} C_0^{1-i} \max_{\|\underline{C}\|_1 = 1} \{C(i,m)\} \right)^{-1}. \quad (53)$$

Paired with the symmetry of the multinominal $C(i, m)$, we apply the Lagrange multiplier optimization and obtain that

$$C(i, m) \leq \binom{m}{i} m^{-i}, \quad (54)$$

where the equality is achieved when having $C_i = m^{-1}$ for all $i \in [m]$. Then, we have

$$\min_{\|\underline{C}\|_1 = 1} \tau^* = \left( 1 - \left( 1 + (mC_0)^{-1} \right)^{-m} \right)^{-1}. \quad (55)$$

## APPENDIX C
## PROOF OF THEOREM 3

With $\mathcal{M}_{\text{local}}$, from Theorem 1, we can easily obtain that $\tau_n^* = (n+1) \cdot (mC)^{-1}$. Applying a change of variables, i.e., $v_n^* = C\tau_n^*$, we can equivalently write the latency above as $v_n^* = m^{-1}(n+1)$, which is treated as the normalized latency with $\mathcal{M}_{\text{local}}$.

With $\mathcal{M}_{\text{global}}$, from Theorem 2, we obtain that

$$\tau_k^* = \frac{\left(C^{-1} + \tau_{k-1}^*\right)\left(C\left(C^{-1} + \tau_{k-1}^*\right)\right)^{m-1}}{1 + \sum_{i=2}^{m}\left(C\left(C^{-1} + \tau_{k-1}^*\right)\right)^{i-1}} \\
= \frac{\left(C\left(C^{-1} + \tau_{k-1}^*\right)\right)^m}{C\sum_{i=0}^{m-1}\left(C\left(C^{-1} + \tau_{k-1}^*\right)\right)^i} = \frac{\tau_{k-1}^*}{1 - \left(1 + C\tau_{k-1}^*\right)^{-m}}, \quad (56)$$

associated with the initial condition $\tau_0^* = m^{-1}C^{-1}$. Applying a change of variables, i.e., $u_k^* = C\tau_k^*$ for all $k \in [n]$, we equivalently have the recursive expression as

$$u_k^* = \left(1 - \left(1 + u_{k-1}^*\right)^{-m}\right)^{-1} u_{k-1}^* \quad (57)$$

with $u_0^* = m^{-1}$, which is treated as the normalized latency with $\mathcal{M}_{\text{global}}$, likewise.

According to the definition of $\rho(n, m)$, we can obtain that

$$\rho(n, m) = \frac{v_n^*}{u_n^*} = \frac{n+1}{m}\left(u_n^*\right)^{-1}. \quad (58)$$

## APPENDIX D
## PROOF OF THEOREM 4

For all $k \in \{0\} \cup [n]$, following the expression of $u_k^*$ used in Theorem 3, we define

$$z_k \triangleq \frac{\bar{\rho}(k)}{k+1} = \lim_{m \to \infty}\left(mu_k^*\right)^{-1}. \quad (59)$$

When $k \leq 1$, we can easily obtain that $z_0 = 1$ and

$$z_1 = \lim_{m \to \infty}\left(m\left(1 - \left(1 + m^{-1}\right)^{-m}\right)^{-1} m^{-1}\right)^{-1} = 1 - e^{-1} \quad (60)$$

For any given $k \geq 2$, since it is known that $z_{k-1}$ is finite and $\lim_{m \to \infty} m^{-1} z_{k-1} = 0$, we have

$$z_k = \lim_{m \to \infty} m^{-1}\left(1 - \left(1 + u_{k-1}^*\right)^{-m}\right)\left(u_{k-1}^*\right)^{-1} \\
= \lim_{m \to \infty}\left(1 - \left(1 + \lim_{m \to \infty}\frac{mu_{k-1}^*}{m}\right)^{-m}\right)\lim_{m \to \infty}\left(mu_{k-1}^*\right)^{-1} \\
= \lim_{m \to \infty}\left(1 - \left(1 + \frac{z_{k-1}^{-1}}{m}\right)^{-m}\right) z_{k-1} = \left(1 - e^{-z_{k-1}^{-1}}\right) z_{k-1}. \quad (61)$$

Thus, we can recursively obtain $z_n$ as $z_n = \left(1 - e^{-z_{n-1}^{-1}}\right) z_{n-1}$, associated with initial condition $z_0 = 1$. Recovering $\bar{\rho}(k)$ in terms of $z_k$, we obtain that

$$\bar{\rho}(n) = \frac{n+1}{n}\left(1 - \exp\left(-\frac{n}{\bar{\rho}(n-1)}\right)\right)\bar{\rho}(n-1), \quad (62)$$

with initial condition $\bar{\rho}(0) = 1$.

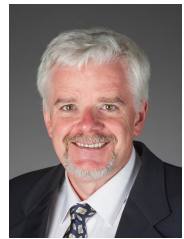

**Martin Haenggi** (S'95-M'99-SM'04-F'14) received the Dipl.-Ing. (M.Sc.) and Dr.sc.techn. (Ph.D.) degrees in electrical engineering from the Swiss Federal Institute of Technology in Zurich (ETH) in 1995 and 1999, respectively. Currently he is the Freimann Professor of Electrical Engineering and a Concurrent Professor of Applied and Computational Mathematics and Statistics at the University of Notre Dame, Indiana, USA. In 2007-2008, he was a visiting professor at the University of California at San Diego, and in 2014-2015 he was an Invited Professor at EPFL, Switzerland. He is a co-author of the monographs "Interference in Large Wireless Networks" (NOW Publishers, 2009) and Stochastic Geometry Analysis of Cellular Networks (Cambridge University Press, 2018) and the author of the textbook "Stochastic Geometry for Wireless Networks" (Cambridge, 2012), and he published 14 single-author journal articles. His scientific interests lie in networking and wireless communications, with an emphasis on cellular, amorphous, ad hoc (including D2D and M2M), cognitive, and vehicular networks. He served as an Associate Editor of the Elsevier Journal of Ad Hoc Networks, the IEEE Transactions on Mobile Computing (TMC), the ACM Transactions on Sensor Networks, as a Guest Editor for the IEEE Journal on Selected Areas in Communications, the IEEE Transactions on Vehicular Technology, and the EURASIP Journal on Wireless Communications and Networking, as a Steering Committee member of the TMC, and as the Chair of the Executive Editorial Committee of the IEEE Transactions on Wireless Communications (TWC). Currently he is the Editor-in-Chief of the TWC. He also served as a Distinguished Lecturer for the IEEE Circuits and Systems Society, as a TPC Co-chair of the Communication Theory Symposium of the 2012 IEEE International Conference on Communications (ICC'12), of the 2014 International Conference on Wireless Communications and Signal Processing (WCSP'14), and the 2016 International Symposium on Wireless Personal Multimedia Communications (WPMC'16). For both his M.Sc. and Ph.D. theses, he was awarded the ETH medal. He also received a CAREER award from the U.S. National Science Foundation in 2005 and three awards from the IEEE Communications Society, the 2010 Best Tutorial Paper award, the 2017 Stephen O. Rice Prize paper award, and the 2017 Best Survey paper award, and he is a 2017 Clarivate Analytics Highly Cited Researcher.

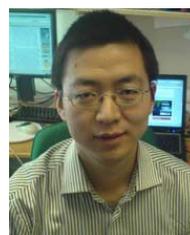

**Ming Xiao** (S'2002-M'2007-SM'2012) received Bachelor and Master degrees in Engineering from the University of Electronic Science and Technology of China, Chengdu in 1997 and 2002, respectively. He received Ph.D degree from Chalmers University of technology, Sweden in November 2007. From 1997 to 1999, he worked as a network and software engineer in ChinaTelecom. From 2000 to 2002, he also held a position in the SiChuan communications administration. From November 2007 to now, he has been in the department of information science and engineering, school of electrical engineering and computer science, Royal Institute of Technology, Sweden, where he is currently an Associate Professor. Since 2012, he has been an Associate Editor for IEEE Transactions on Communications, IEEE Communications Letters (Senior Editor Since Jan. 2015) and IEEE Wireless Communications Letters (2012-2016). He was the lead Guest Editor for IEEE JSAC Special issue on "Millimeter Wave Communications for future mobile networks" in 2017.

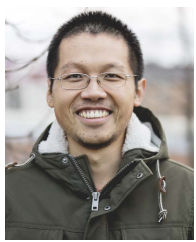

**Guang Yang** received his B.E degree in Communication Engineering from University of Electronic Science and Technology of China (UESTC), Chengdu, China in 2010, and from 2010 to 2012 he participated in the joint Master-PhD program in National Key Laboratory of Science and Technology on Communications at UESTC. He joined the Department of Information Science and Engineering at the School of Electrical Engineering and Computer Science, the Royal Institute of Technology (KTH), Stockholm, Sweden, as a Ph.D. student since September of 2013. He was a visiting student at University of Notre Dame, US, in 2017.